\documentclass[12pt]{article}

\usepackage{scicite}

\usepackage{times}

\usepackage{amsmath}
\usepackage{amsfonts}
\usepackage{amssymb}
\usepackage{graphicx}
\usepackage[colorlinks]{hyperref}
\usepackage{caption}
\newenvironment{sciabstract}{%
\begin{quote} \bf}
{\end{quote}}

\title{Abrupt transitions in collaborative social networks}

\author
{Jingfang Fan,${}^{1,2\ddagger}$ Jun Meng,${}^{1, 3\ddagger}$ Yimin Ding,${}^{4}$ Guangle Du,${}^{3}$ Daqing Li,${}^{5,6}$ \\ Reuven Cohen,${}^{7}$ Xiaosong Chen,${}^{2,8\ast}$
Fangfu Ye,${}^{3,8\ast}$  Shlomo Havlin${}^{1\ast}$\\
\\
\normalsize{${}^{1}$Department of Physics, Bar Ilan University, Ramat Gan 52900, Israel} \\
\normalsize{${}^{2}$CAS Key Lab of Theoretical Physics, Institute of Theoretical Physics}, \\ \normalsize{Chinese Academy of Sciences, Beijing 100190, China} \\
\normalsize{${}^{3}$Beijing National Laboratory for Condensed Matter Physics \& CAS Key Lab of Soft Matter Physics}, \\  \normalsize{Institute of Physics, Chinese Academy of Sciences, Beijing 100190, China}\\
\normalsize{${}^{4}$Faculty of Physics and Electronics, Hubei University, Wuhan 43006, China} \\
\normalsize{${}^{5}$School of Reliability and Systems Engineering, Beihang University, Beijing 100191, China} \\
\normalsize{${}^{6}$Science and Technology on Reliability and Environmental Engineering Laboratory}, \\ \normalsize{Beijing 100191, China} \\
\normalsize{${}^{7}$Department of Mathematics, Bar Ilan University, Ramat Gan 52900, Israel} \\
\normalsize{${}^{8}$School of Physical Sciences, University of Chinese Academy of Sciences, Beijing 100049, China} \\
}

\date{}

\begin{document} 


\baselineskip24pt


\maketitle


\begin{sciabstract}
Despite the wide use of networks as a versatile tool for 
exploring complex social systems, little is known about how 
to detect and forecast abrupt changes in social systems. 
In this report, we develop an early warning approach based on network properties to detect such changes. By analysing three collaborative social networks---one co-stardom, one patent and one scientific collaborative network, we discover that abrupt transitions inherent in these networks can serve as a good early warning signal, indicating, respectively, the dissolution of the Soviet Union, the emergence of the ``soft matter" research field, and the merging of two scientific communities. We then develop a clique growth model that 
explains the universal properties of these real networks
and find that they belong to a new universality class, described by the Gumbel distribution. 
%
%
\end{sciabstract}

\section*{Introduction}

Networks have demonstrated their great potential as versatile tools for 
exploring the dynamical and structural properties
of complex systems from a wide variety of disciplines in  physics,
biology, neuroscience and social science \cite{watts_collective_1998,barabasi_emergence_1999,newman2010networks}. They are used to predict the evolution of a scientist's impact \cite{sinatra_quantifying_2016}, quantifyand forecast disease epidemics \cite{brockmann_hidden_2013},  study the role of the airline transportation network in the prediction and predictability of global epidemics\cite{colizza2006role},
forecast climate extreme events \cite{ludescher_very_2014,boers_prediction_2014}, and more \cite{cohen2010complex}. Despite their widespread use, there is still a lack of a quantitative early warning system to detect the approaching of abrupt global social changes.  Global social changes refer to a global  transformation of behaviors, culture, and social structure over time \cite{scheffer2009critical}.  Abrupt global change may imply risks of unwanted collapses, such as political conflicts (e.g., the recent violence in Charlottesville in Virginia, USA) and/or financial crises, or opportunities for positive changes, such as culture integration and/or technology revolution. These changes are usually associated with so-called tipping points, critical thresholds at which a tiny perturbation can qualitatively alter the state or development of a system \cite{lenton_tipping_2008,scheffer_anticipating_2012}.
The capacity to predict such tipping points and thus identify the corresponding risks and opportunities would be greatly boosted if one could reveal the fundamental architectural changes that might occur in social networks. 

In this study, we develop an early warning approach based on network and percolation theories \cite{essam_percolation_1980} to detect tipping points and use it to forecast abrupt social changes. 
We investigate three social collaborative networks----a co-stardom network, a patent network, and a scientific collaborative network, which represent, respectively, cultural/political, technological, and scientific social systems. 
In the co-stardom network, we treat actors as nodes and films as cliques, i.e., sets of nodes that are all connected to each other \cite{bollobas2001random}. We analyze the evolution of this network using our method and find that the network undergoes an abrupt transition with the addition of a single critical film, which was released in 1988. This transition can be viewed as a warning signal for the dissolution of the Soviet Union in 1991, as can be seen clearly in the film networks. In the patent network, the inventors are nodes and patents are cliques (of several inventors); we also find an abrupt transition in 1989, which can be interpreted as an indicator of the emergence of a new research field----soft matter----and the awarding of the 1991 Nobel Prize in physics to P. G. de Gennes for his contributions to the studies of liquid crystals and polymers. Similarly, in the scientific collaborative network, in which the scientists/authors are nodes and articles are cliques (of several authors), there also exists an abrupt transition, which forecasts the merging of two scientific communities.  In order to better understand and characterize the evolution of  these networks, 
we develop a clique growth model, based on the real data, i.e., the clique-size distribution and the activity distribution of nodes are taken from these real networks. This model not only reproduces the results observed in the three real social networks but also shows a novel universality class which is different from the common clique model constructed by growing the networks according to the random selection rule of links as in the classical Erd\H os-R\'enyi (ER) networks \cite{erdos1960evolution}.

\section*{Precursors of abrupt social changes}

Significant social changes are usually associated with integration and collaboration of distinct social communities.  Clique networks have been reported to be a useful tool to 
characterize and quantify the evolution of social communities \cite{palla_uncovering_2005}.  A key concept in clique networks is 	a ``component", 
with an $(m,l)$ value defined to be the nodes belonging to a series of overlapping $m$-cliques, 
where overlap means that two 
$m$-cliques share at least $l$-nodes \cite{derenyi_clique_2005} 
and $m$ represents the number of nodes in the individual cliques. 
We are interested in understanding the growth process of  social clique networks. As more and more cliques are added, the size of the network increases, and its structure changes accordingly. 
We monitor the changes of the size $S$ of the largest $(m,l)$ component during the network growth process,
because we hypothesize that an abrupt jump 
in $S$ could be associated to an abrupt global social change \cite{shirky2011political,kaplan2010users}. We focus here on 
the largest $(3,2)$ component, as it turns out to be  the simplest component that shows abrupt jumps.

\begin{figure}
\begin{centering}
\includegraphics[width=1.0\linewidth]{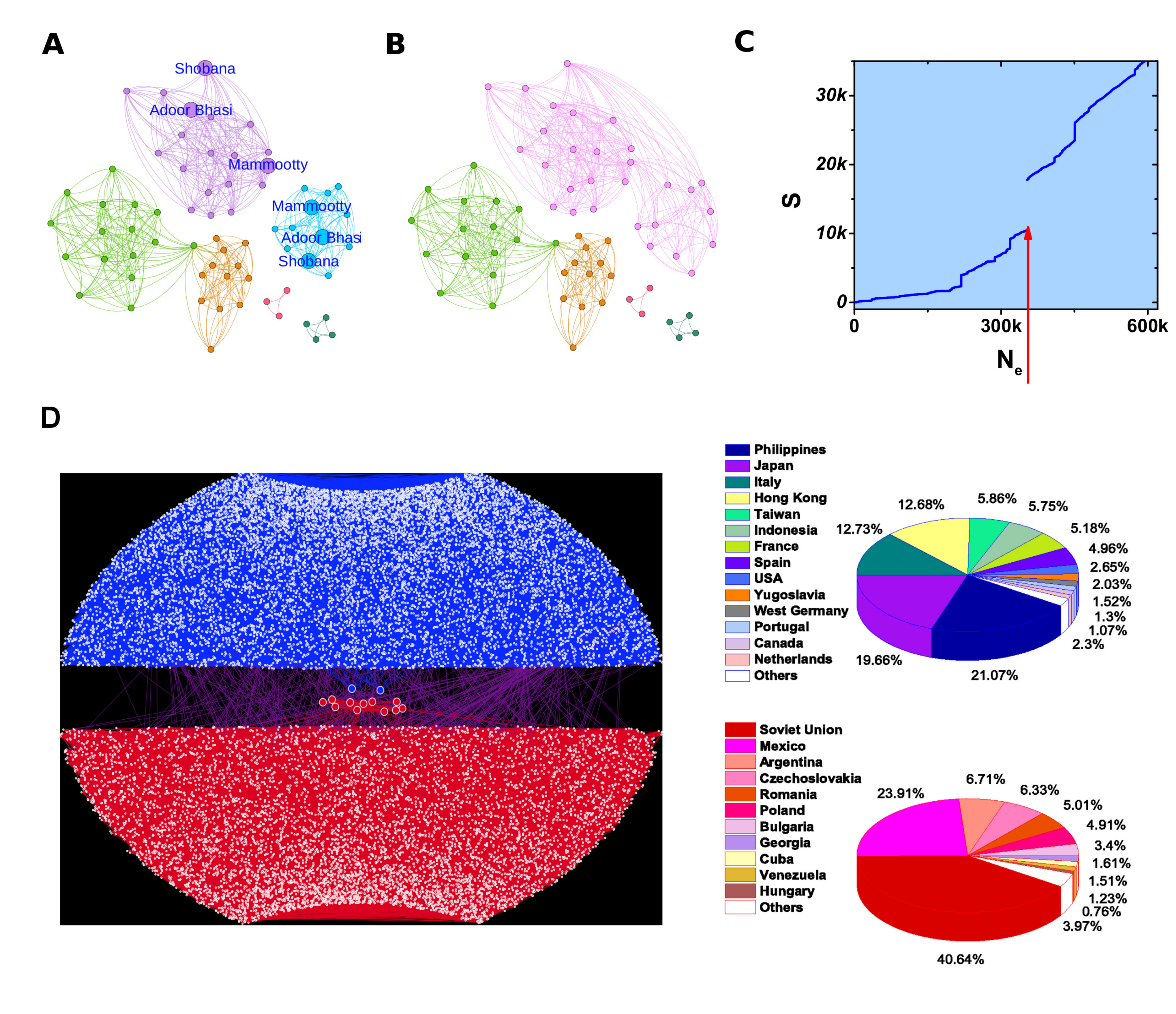}
\caption{\label{Fig:1}  Illustration of the evolution of the co-stardom network. (A) The nodes (actors) with a same color belong to a same film and are thus connected to each other; (B) merge of two cliques that share three actors [represented by the large nodes in (A)]; (C)  relation between the size $S$ of the largest $(3,2)$ component and the total number of the network edges $N_e$; (D) merge of two giant $(3,2)$ components at the abrupt transition point, where the thirteen large nodes in the center of the left figure represent the critical clique/film and the two pie charts in the right give, respectively, 
the distributions of the countries releasing the films in the two giant components. Note, in the top are mainly Western countries while in the bottom are Eastern countries.}
\end{centering}
\end{figure}

We start with investigating the co-stardom network, which is formed by all the films released between 1985 and 2015 in the IMDb dataset ~\cite{IMDb}. The network is constructed by adding film by film starting from the first film in 1985 according to their release time. Given that the films usually had different numbers of actors, the corresponding sizes of the cliques are added at each step are usually different from each other. 
Figs.~\ref{Fig:1} (A) and (B) illustrate the growth process of the network and the formation of $(3,2)$ components. As shown in Figs.~\ref{Fig:1} (C), the size change of the largest $(3,2)$ component shows an abrupt jump with the addition of a critical clique, a film with 13 actors released by Venezuela in 1988. The origin of this jump is that this film merges two large $(3,2)$ components----one composed of films mainly contributed by Western countries and the other of films by Eastern social states [see Figs.~\ref{Fig:1} (D)]. 
Indeed, this merge in 1988 reflected the rapid increase of the communication between the two distinct political camps. Note that 1988 was also the year when Moscow lost control \cite{Soviet_Union}, and three years later the Soviet Union dissolved.

To demonstrate the robustness of our method, we apply it also to two other real networks----one patent network and one scientific collaborative network. The patent network is formed by all U.S. utility patents published between 1975 and 1990 from the Google patent dataset ~\cite{Google}. We carried out the same analysis, and observed an abrupt size jump of the largest $(3,2)$ component when a critical patent, published in 1989 (titled `Nematic liquid-crystalline phases'), was added. This is because this patent links together two large components, which correspond, respectively, to the Physics and Chemistry classifications [see Fig.~S2 in the Supplementary Materials~\cite{SI}]. This merge of `Physics' and 'Chemistry' had a consequence: the emergence of a new interdisciplinary research field----soft matter, which resulted from interactions between physicists and chemists and formally came about after P. G. de Gennes' Nobel Prize speech in 1991 \cite{de1992soft}.  
The scientific collaborative network we analysed is composed of physics articles published within last 10 years (2006-2016) with topic `networks', which were collected from the Web of Science database \cite{webofscience}. The same analysis yields an abrupt jump and identifies a critical article (published in 2014, see Table.~S1 in the Supplementary Materials~\cite{SI}), which merges the component of scientists who focus on non-quantum phenomena with that of scientists who mainly study quantum phenomena [see Fig.~ S3]. This merging implies the beginning of strong collaboration between non-quantum physicists and quantum physicists in the field of networks. To clearly demonstrate the results of this merging, we use `quantum network' as a keyword to search these articles in the `networks' topic. Such a keyword reflects the influence of networks studies on physicists studying quantum phenomena. We find that the number of papers containing this quantum-network keyword before the addition of this critical article is 0 (out of 33281 articles) and that this number became 25 (out of 5861) after the merging of these two communities. Thus, the abrupt jump in the scientific collaborative network is useful as a forecast tool for scientific integration.

\section*{Clique Growth Model and Universality Class}

To better characterize and understand the aforementioned social networks,  we introduce a clique  growth model (CGM). At each time step, $\tilde{m}$ nodes from given $N$ total nodes  
are randomly chosen and connected to each other. Thus, at each step the growth process yields at least one $\tilde{m}$-clique. Fig.~S4 illustrates a network formed by $3$-cliques. 
Considering that the cliques in the real networks do not have a fixed size, we assign 
a distribution, $p(\tilde{m})$, to the number of nodes chosen at each step. The simplest distribution is, for example, $p(\tilde{m})=\delta_{\tilde{m}3}$, which means only three nodes are chosen at each step during the network growth. Another distribution we need is the distribution of number of times $w$ an individual node appear in all $\tilde{m}$-cliques, which we call ``activity distribution" $\phi(w)$ 
[see Sec.~I of the Supplementary Materials for more details of the definitions of $p(\tilde{m})$ and $\phi(w)$].
In principle, these two distributions can take any form; however, to mimic the real networks, we start with their forms being the same as their counterparts in real data, as shown in Fig.~S1, and we call the model constructed this way ``Model A".


To better understand the origin of the jump and  whether the abrupt jumps reported above corresponds to continuous or discontinuous percolations in the limit of an infinitely large system, we test our model where we can control the size of the network. For this we calculate in our clique growth model the fraction of the largest size change of the largest $(m,l)$ component: \begin{equation}
\Delta \equiv \frac{1}{N}\max\left[S_2-S_1,\cdots,S_{t+1}-S_t,\cdots\right],
\label{eq1}
\end{equation}
where $S_t$ represents the size of the largest $(m,l)$ component at step $t$ and $N$ is the total number of nodes. 
If $\Delta$ approaches zero as $N\rightarrow\infty$, the corresponding $(m,l)$ component undergoes a continuous percolation; otherwise, the corresponding percolation is discontinuous.

To consider the randomness associated with the growth process of CGM, we run 1000 realizations for a given set of $[N,p(\tilde{m}),\psi(w)]$, obtaining one value of $\Delta$ for each realization. The resulted average $\bar{\Delta}$ and standard deviation $\chi_\Delta= \left<[\Delta-\bar{\Delta}]^2\right>^{1/2}$  is assumed to decay with the system size $N$, respectively, as $\bar{\Delta}(N) \propto  N^{-\beta_{1}}$ and $\chi_{\Delta} \propto  N^{-\beta_2}$, where $\beta_{1}$ and $\beta_{2}$ are two critical exponents characterizing the universality class of percolation in this growth system. Different clique percolation models with the same critical exponents belong to the same universality class. If $\beta_1$ (or $\beta_2$) is zero, the corresponding percolation is discontinuous.

\begin{figure}
\begin{centering}
\includegraphics[width=1.0\linewidth]{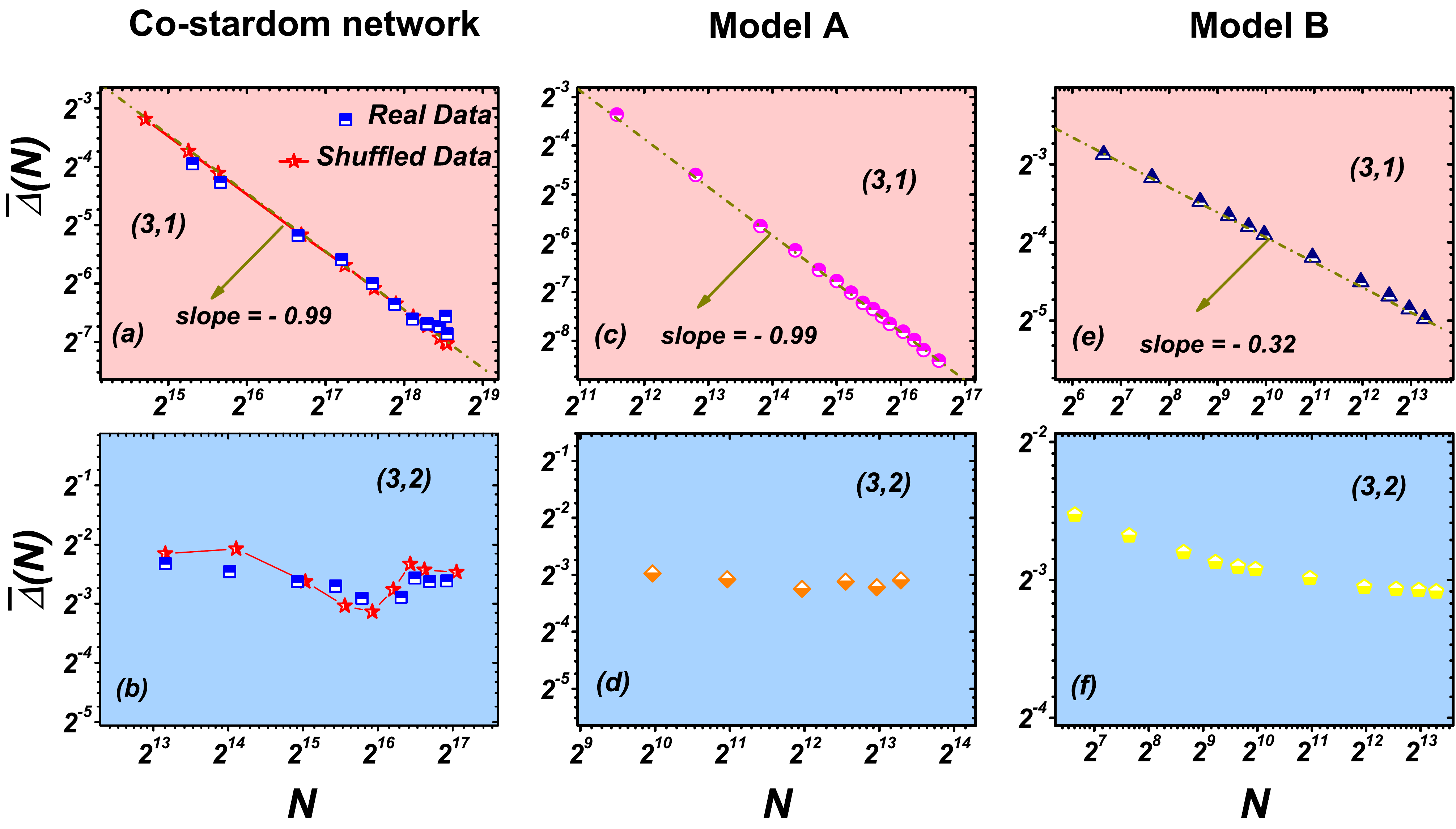}
\caption{\label{Fig:2}  Log-log plots of the average size change $\bar{\Delta}(N)$ of the $(3,1)$ and $(3,2)$ components 
versus the network size $N$ for the co-stardom network [(a) and (b)], Model A [(c) and (d)], and Model B [(e) and (f)].}
\end{centering}
\end{figure}

We first present the numerical results for $\bar{\Delta}(N)$ (the results for $\chi_\Delta$ are given in the Supplementary Materials).  We focus on the model based on the co-stardom network (with the results of the other two cases given in the Supplementary Materials), viz., we assume $p(\tilde{m})$ and $\psi(w)$ have the same forms as those found in the real co-stardom network [see Fig.~S1].  As seen in Figs.~\ref{Fig:2} (c) and (d), when $N\rightarrow\infty$,  $\bar{\Delta}$ of the $(3,2)$ component tends to a finite value; but that of the $(3,1)$ component tends to zero, with $\beta_1\approx 0.99$.  These results agree with those of the corresponding real data [see Figs.~\ref{Fig:2} (a) and (b)]: the percolation of the $(3,2)$ component is discontinuous while that of the $(3,1)$ component is continuous. This difference of percolation behaviours arises because
it requires a significant collaboration (more than one common actor in two movies) to obtain a giant $(3,2)$ component.  

To clarify the effect of the release time of the films (other than providing timing information for predicting social events) on the percolation process, 
we randomly shuffled the timing of the films in the real data 1000 times, yielding 1000 realizations of networks. Note that the shuffling by itself does not change $p(\tilde{m})$ and $\psi(w)$ (see the Supplementary Materials for details). The results of the shuffled real data are also shown in Figs.~\ref{Fig:2} (a) and (b). We can clearly see the continuity behaviours of the percolation and the value of $\beta_1$ does not depend on the timing of the films, i.e., the shuffling does not change the universality class of the percolations. 



The aforementioned results not only confirm that the abrupt transitions observed in the real collaborative social networks correspond to the discontinuous transitions of the $(3,2)$ components of the CGM but also suggest that percolation in these networks belong to a new universality class, because 
the critical exponent $\beta_1\approx 0.99$ 
is different from the one in the random ER and other related models of $1/3$\cite{nagler_impact_2011}. To investigate the origin of this difference, 
we construct
another model, which we call ``Model B", where $p(\tilde{m})=\delta_{\tilde{m}3}$ and $\phi(w)$ has a uniform random distribution, viz., each node has an equal probability of being chosen at each step during the network growth process. In such a model, as shown in
Figs. \ref{Fig:2} (e) and (f), the $\bar{\Delta}$ of the $(3,2)$ component still tends to a non-zero constant as $N\rightarrow\infty$; however, the exponent $\beta_1$ in this model for the $(3,1)$ component is now close to $1/3$. 
We also try other forms of $p(\tilde{m})$ and $\phi(w)$, whose results are given in Fig.~S5 of the Supplementary Materials, and find that the value of $\beta_1$ or the universality class of the corresponding percolations does not depend on the form of $p(\tilde{m})$ but changes with the form of $\phi(w)$.  

\begin{figure}
\begin{centering}
\includegraphics[width=1.0\linewidth]{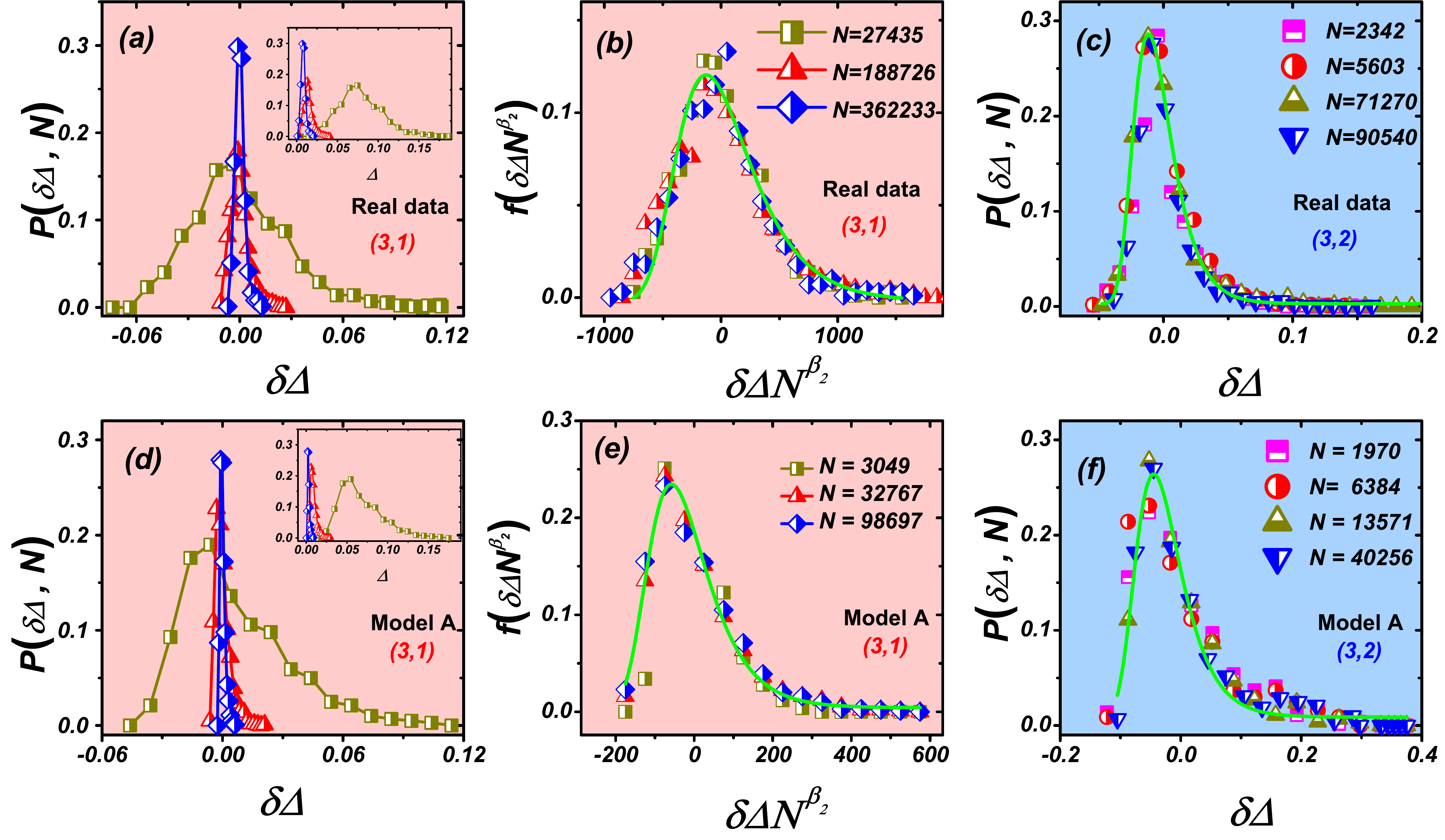}
\caption{\label{Fig:3} Finite-size scaling functions  $P(\delta \Delta,N)$ and their universal functions $f(\delta \Delta N^{\beta_2})$
for real shuffled data of the co-stardom network [(a), (b), and (c)] and the clique growth Model A [(d), (e), and (f)].  
The green lines in the figures are fitting lines obtained from Eq.~(\ref{eq11}). 
The insets  in (a) and (d) are the distribution $P$ expressed as function of $\Delta$ rather than $\delta \Delta$.}
\end{centering}
\end{figure}

To further reveal the difference between the continuous (in $(3,1)$) and discontinuous (in $(3,2)$) transitions, we  proceed to investigate the distribution of the fluctuation $\delta \Delta=\Delta-\bar{\Delta}$,
which is expected to take the following finite-size scaling form \cite{privman_universal_1984,fan2014general},
\begin{equation}
P(\delta \Delta,N) =  N^{\beta_{2}}f(\delta \Delta N^{\beta_2}),
\label{eq10}
\end{equation}
where $f(\cdot)$ is a universal function and $\beta_2$ is the critical exponent of the scaling behaviours of the standard deviation $\chi_{\Delta}$.
Given that $\Delta$ is the maximum value of the changes of $S(t)$ resulting from the various random realizations, the distribution of $\delta \Delta$ is expected to follow the Gumbel extreme value distribution, which means the universal function in Eq.~(\ref{eq10}) can be written as \cite{gumbel2012statistics},
\begin{equation}
f (\delta \Delta N^{\beta_2})= f_{0} + A*e^{-e^{-z}-z},  \textrm{\ \ \ \ with \ \ } z = (\delta \Delta N^{\beta_2}-B)/\omega,
\label{eq11}
\end{equation}
where $f_{0}, A, B$ and $\omega$ are four constants 
and the value of the critical exponent $\beta_2$ can be obtained by numerically fitting data of $\chi_{\Delta}$ [see Fig.~S7]. Indeed, the results given in Figs.~\ref{Fig:3} support that the fluctuations of $\Delta$ in both the shuffled real data and the CGM obey a finite-size scaling form of the Gumbel distribution, with 
%
%
$\beta_2\approx 0.99$ for the $(3,1)$ component and $\beta_2=0$ for the $(3,2)$ component.
%
%
These two values of $\beta_2$ echo the results presented above.



\section*{Conclusion}
In conclusion, we analysed three collaborative social networks and found that the discontinuous percolation phenomena found in   the $(3,2)$ clique components in these networks can serve as an early warning signal to predict abrupt social changes.  The clique growth model we developed based on real data activity distribution reveals that these real networks share a universal fluctuation distribution given by the Gumbel extreme function and represent a new universality class with novel critical exponents determined by the activity distribution of individual nodes. 
Our approach is geared towards identifying the critical events that herald revolutions in human societies, and through its use we have identified three events that  have influenced specific global transition instances in political, technological, and scientific societies. Our method of analysis provides a new perspective on the evolution of social networks, and can potentially be used as a template for the discovery of tipping points in other complex systems.
\bibliography{MyLibrary}

\begin{thebibliography}{10}

\bibitem{watts_collective_1998}
D.~J. Watts, S.~H. Strogatz, {\it Nature\/} {\bf 393}, 440 (1998).

\bibitem{barabasi_emergence_1999}
A.-L. Barab{\'a}si, R.~Albert, {\it Science\/} {\bf 286}, 509 (1999).

\bibitem{newman2010networks}
M.~Newman, {\it Networks: an introduction\/} (Oxford university press, 2010).

\bibitem{sinatra_quantifying_2016}
R.~Sinatra, D.~Wang, P.~Deville, C.~Song, A.-L. Barab{\'a}si, {\it Science\/}
  {\bf 354}, 5239 (2016).

\bibitem{brockmann_hidden_2013}
D.~Brockmann, D.~Helbing, {\it Science\/} {\bf 342}, 1337 (2013).

\bibitem{colizza2006role}
V.~Colizza, A.~Barrat, M.~Barth{\'e}lemy, A.~Vespignani, {\it Proceedings of
  the National Academy of Sciences of the United States of America\/} {\bf
  103}, 2015 (2006).

\bibitem{ludescher_very_2014}
J.~Ludescher, {\it et~al.\/}, {\it Proceedings of the National Academy of
  Sciences\/} {\bf 111}, 2064 (2014).

\bibitem{boers_prediction_2014}
N.~Boers, {\it et~al.\/}, {\it Nature Communications\/} {\bf 5}, 5199 (2014).

\bibitem{cohen2010complex}
R.~Cohen, S.~Havlin, {\it Complex networks: structure, robustness and
  function\/} (Cambridge university press, 2010).

\bibitem{scheffer2009critical}
M.~Scheffer, {\it Critical transitions in nature and society\/} (Princeton
  University Press, 2009).

\bibitem{lenton_tipping_2008}
T.~M. Lenton, {\it et~al.\/}, {\it Proceedings of the National Academy of
  Sciences\/} {\bf 105}, 1786 (2008).

\bibitem{scheffer_anticipating_2012}
M.~Scheffer, {\it et~al.\/}, {\it Science\/} {\bf 338}, 344 (2012).

\bibitem{essam_percolation_1980}
J.~W. Essam, {\it Reports on Progress in Physics\/} {\bf 43}, 833 (1980).

\bibitem{bollobas2001random}
B.~Bollob{\'a}s, {\it Random graphs\/} (Cambridge University Press, 2001).

\bibitem{erdos1960evolution}
P.~Erd{\H{o}}s, A.~R{\'e}nyi, {\it Publ. Math. Inst. Hung. Acad. Sci\/} {\bf
  5}, 17 (1960).

\bibitem{palla_uncovering_2005}
G.~Palla, I.~Der{\'e}nyi, I.~Farkas, T.~Vicsek, {\it Nature\/} {\bf 435}, 814
  (2005).

\bibitem{derenyi_clique_2005}
I.~Der{\'e}nyi, G.~Palla, T.~Vicsek, {\it Physical Review Letters\/} {\bf 94},
  160202 (2005).

\bibitem{shirky2011political}
C.~Shirky, {\it Foreign affairs\/} pp. 28--41 (2011).

\bibitem{kaplan2010users}
A.~M. Kaplan, M.~Haenlein, {\it Business horizons\/} {\bf 53}, 59 (2010).

\bibitem{IMDb}
{\it http://www.imdb.com/\/}.

\bibitem{Soviet_Union}
{\it Government in the Soviet Union: Gorbachev's Proposal for Change\/} (New
  York Times., 1988).

\bibitem{Google}
{\it https://patents.google.com/\/}.

\bibitem{SI}
{\it Supplementary materials\/}.

\bibitem{de1992soft}
P.-G. de~Gennes, {\it Angewandte Chemie International Edition\/} {\bf 31}, 842
  (1992).

\bibitem{webofscience}
{\it https://www.webofknowledge.com/\/}.

\bibitem{nagler_impact_2011}
J.~Nagler, A.~Levina, M.~Timme, {\it Nature Physics\/} {\bf 7}, 265 (2011).

\bibitem{privman_universal_1984}
V.~Privman, M.~E. Fisher, {\it Physical Review B\/} {\bf 30}, 322 (1984).

\bibitem{fan2014general}
J.~Fan, X.~Chen, {\it EPL (Europhysics Letters)\/} {\bf 107}, 28005 (2014).

\bibitem{gumbel2012statistics}
E.~J. Gumbel, {\it Statistics of extremes\/} (Courier Corporation, 2012).

\end{thebibliography}

\bibliographystyle{Science}

\section*{Acknowledgments}
We thank James Farrell for helpful discussions. This work was supported by the Hundred-Talent Program of the Chinese Academy of Sciences (CAS), the Key Research Program of Frontier Sciences of CAS [Grant No. QYZDB-SSW-SYS003], the National Natural Science Foundation of China [Grant 1121403], the fellowship program of the Planning and Budgeting Committee of the Council for Higher Education of Israel, the  Israel Ministry of Science and Technology (MOST) with the Italy Ministry of Foreign Affairs, MOST with the Japan Science and Technology Agency, the BSF-NSF foundation, the Israel Science Foundation, ONR and DTRA. 

\section*{Supplementary materials}
Materials and Methods\\
Supplementary Text\\
Figs. S1 to S8\\
Tables S1\\
References \textit{(30-33)}


\clearpage

\end{document}